\tolerance=500
%
\catcode`@=11 
%
%
%
\font\twentyrm=cmr10 scaled\magstep4
\font\seventeenrm=cmr10 scaled\magstep3
\font\fourteenrm=cmr10 scaled\magstep2
\font\twelverm=cmr10 scaled\magstep1
\font\ninerm=cmr9
\font\eightrm=cmr8
\font\sixrm=cmr6
\font\twentybf=cmbx10 scaled\magstep4
\font\seventeenbf=cmbx10 scaled\magstep3
\font\fourteenbf=cmbx10 scaled\magstep2
\font\twelvebf=cmbx10 scaled\magstep1
\font\ninebf=cmbx9
\font\eightbf=cmbx8
\font\sixbf=cmbx6
\font\twentyi=cmmi10 scaled\magstep4        \skewchar\twentyi='177
\font\seventeeni=cmmi10 scaled\magstep3     \skewchar\seventeeni='177
\font\fourteeni=cmmi10 scaled\magstep2      \skewchar\fourteeni='177
\font\twelvei=cmmi10 scaled\magstep1        \skewchar\twelvei='177
\font\ninei=cmmi9                           \skewchar\ninei='177
\font\eighti=cmmi8                          \skewchar\eighti='177
\font\sixi=cmmi6                            \skewchar\sixi='177
\font\twentysy=cmsy10 scaled\magstep4       \skewchar\twentysy='60
\font\seventeensy=cmsy10 scaled\magstep3    \skewchar\seventeensy='60
\font\fourteensy=cmsy10 scaled\magstep2     \skewchar\fourteensy='60
\font\twelvesy=cmsy10 scaled\magstep1       \skewchar\twelvesy='60
\font\ninesy=cmsy9                          \skewchar\ninesy='60
\font\eightsy=cmsy8                         \skewchar\eightsy='60
\font\sixsy=cmsy6                           \skewchar\sixsy='60

\font\twentyex=cmex10 scaled\magstep4
\font\seventeenex=cmex10 scaled\magstep3
\font\fourteenex=cmex10 scaled\magstep2
\font\twelveex=cmex10 scaled\magstep1

\font\twentysl=cmsl10 scaled\magstep4
\font\seventeensl=cmsl10 scaled\magstep3
\font\fourteensl=cmsl10 scaled\magstep2
\font\twelvesl=cmsl10 scaled\magstep1
\font\ninesl=cmsl9
\font\eightsl=cmsl8
\font\twentyit=cmti10 scaled\magstep4
\font\seventeenit=cmti10 scaled\magstep3
\font\fourteenit=cmti10 scaled\magstep2
\font\twelveit=cmti10 scaled\magstep1
\font\nineit=cmti9
\font\eightit=cmti8
\font\twentytt=cmtt10 scaled\magstep4
\font\seventeentt=cmtt10 scaled\magstep3
\font\fourteentt=cmtt10 scaled\magstep2
\font\twelvett=cmtt10 scaled\magstep1
\font\ninett=cmtt9
\font\eighttt=cmtt8
\font\twentycp=cmcsc10 scaled\magstep4
\font\seventeencp=cmcsc10 scaled\magstep3
\font\fourteencp=cmcsc10 scaled\magstep2
\font\twelvecp=cmcsc10 scaled\magstep1
\font\tencp=cmcsc10
\newfam\cpfam
%
%
%
%
\fontdimen16\twentysy=5.40pt \fontdimen17\twentysy=5.40pt
\fontdimen16\seventeensy=4.59pt \fontdimen17\seventeensy=4.59pt
\fontdimen16\fourteensy=3.78pt \fontdimen17\fourteensy=3.78pt
\fontdimen16\twelvesy=3.24pt \fontdimen17\twelvesy=3.24pt
\fontdimen16\tensy=2.70pt \fontdimen17\tensy=2.70pt
\fontdimen16\ninesy=2.43pt \fontdimen17\ninesy=2.43pt
\fontdimen16\eightsy=2.16pt \fontdimen17\eightsy=2.16pt
\fontdimen16\sixsy=1.62pt \fontdimen17\sixsy=1.62pt
%
%
%
\newcount\f@ntkey \f@ntkey=0
\def\samef@nt{\relax\ifcase \f@ntkey \rm \or\oldstyle \or\or
  \or\it \or\sl \or\bf \or\tt \or\caps \fi}
%
%
%
\def\eightpoint{\relax
  \textfont0=\eightrm \scriptfont0=\sixrm \scriptscriptfont0=\fiverm
  \textfont1=\eighti \scriptfont1=\sixi \scriptscriptfont1=\fivei
  \textfont2=\eightsy \scriptfont2=\sixsy \scriptscriptfont2=\fivesy
  \textfont3=\tenex \scriptfont3=\tenex \scriptscriptfont3=\tenex
  \textfont\itfam=\eightit
  \textfont\slfam=\eightsl
  \textfont\bffam=\eightbf \scriptfont\bffam=\sixbf
    \scriptscriptfont\bffam=\fivebf
  \textfont\ttfam=\eighttt
  \textfont\cpfam=\tencp
  \def\rm{\fam0 \eightrm \f@ntkey=0 }
  \def\oldstyle{\fam1 \eighti \f@ntkey=1 }
  \def\it{\fam\itfam \eightit \f@ntkey=4 }
  \def\sl{\fam\slfam \eightsl \f@ntkey=5 }
  \def\bf{\fam\bffam \eightbf \f@ntkey=6 }
  \def\tt{\fam\ttfam \eighttt \f@ntkey=7 }
  \def\caps{\fam\cpfam \tencp \f@ntkey=8 }
  \h@big=6.8\p@{} \h@Big=9.2\p@{} \h@bigg=11.6\p@{} \h@Bigg=14\p@{}
  \setbox\strutbox=\hbox{\vrule height 6.8pt depth 2.8pt width\z@}
  \samef@nt}
\def\ninepoint{\relax
  \textfont0=\ninerm \scriptfont0=\sixrm \scriptscriptfont0=\fiverm
  \textfont1=\ninei \scriptfont1=\sixi \scriptscriptfont1=\fivei
  \textfont2=\ninesy \scriptfont2=\sixsy \scriptscriptfont2=\fivesy
  \textfont3=\tenex \scriptfont3=\tenex \scriptscriptfont3=\tenex
  \textfont\itfam=\nineit
  \textfont\slfam=\ninesl
  \textfont\bffam=\ninebf \scriptfont\bffam=\sixbf
    \scriptscriptfont\bffam=\fivebf
  \textfont\ttfam=\ninett
  \textfont\cpfam=\tencp
  \def\rm{\fam0 \ninerm \f@ntkey=0 }
  \def\oldstyle{\fam1 \ninei \f@ntkey=1 }
  \def\it{\fam\itfam \nineit \f@ntkey=4 }
  \def\sl{\fam\slfam \ninesl \f@ntkey=5 }
  \def\bf{\fam\bffam \ninebf \f@ntkey=6 }
  \def\tt{\fam\ttfam \ninett \f@ntkey=7 }
  \def\caps{\fam\cpfam \tencp \f@ntkey=8 }
  \h@big=7.65\p@{} \h@Big=10.35\p@{} \h@bigg=13.05\p@{} \h@Bigg=15.75\p@{}
  \setbox\strutbox=\hbox{\vrule height 7.65pt depth 3.15pt width\z@}
  \samef@nt}
\def\tenpoint{\relax
  \textfont0=\tenrm \scriptfont0=\sevenrm \scriptscriptfont0=\fiverm
  \textfont1=\teni \scriptfont1=\seveni \scriptscriptfont1=\fivei
  \textfont2=\tensy \scriptfont2=\sevensy \scriptscriptfont2=\fivesy
  \textfont3=\tenex \scriptfont3=\tenex \scriptscriptfont3=\tenex
  \textfont\itfam=\tenit
  \textfont\slfam=\tensl
  \textfont\bffam=\tenbf \scriptfont\bffam=\sevenbf
    \scriptscriptfont\bffam=\fivebf
  \textfont\ttfam=\tentt
  \textfont\cpfam=\tencp
  \def\rm{\fam0 \tenrm \f@ntkey=0 }
  \def\oldstyle{\fam1 \teni \f@ntkey=1 }
  \def\it{\fam\itfam \tenit \f@ntkey=4 }
  \def\sl{\fam\slfam \tensl \f@ntkey=5 }
  \def\bf{\fam\bffam \tenbf \f@ntkey=6 }
  \def\tt{\fam\ttfam \tentt \f@ntkey=7 }
  \def\caps{\fam\cpfam \tencp \f@ntkey=8 }
  \h@big=8.5\p@{} \h@Big=11.5\p@{} \h@bigg=14.5\p@{} \h@Bigg=17.5\p@{}
  \setbox\strutbox=\hbox{\vrule height 8.5pt depth 3.5pt width\z@}
  \samef@nt}
\def\twelvepoint{\relax
  \textfont0=\twelverm \scriptfont0=\ninerm \scriptscriptfont0=\sixrm
  \textfont1=\twelvei \scriptfont1=\ninei \scriptscriptfont1=\sixi
  \textfont2=\twelvesy \scriptfont2=\ninesy \scriptscriptfont2=\sixsy
  \textfont3=\twelveex \scriptfont3=\twelveex \scriptscriptfont3=\twelveex
  \textfont\itfam=\twelveit
  \textfont\slfam=\twelvesl \scriptfont\slfam=\ninesl
  \textfont\bffam=\twelvebf \scriptfont\bffam=\ninebf
    \scriptscriptfont\bffam=\sixbf
  \textfont\ttfam=\twelvett
  \textfont\cpfam=\twelvecp
  \def\rm{\fam0 \twelverm \f@ntkey=0 }
  \def\oldstyle{\fam1 \twelvei \f@ntkey=1 }
  \def\it{\fam\itfam \twelveit \f@ntkey=4 }
  \def\sl{\fam\slfam \twelvesl \f@ntkey=5 }
  \def\bf{\fam\bffam \twelvebf \f@ntkey=6 }
  \def\tt{\fam\ttfam \twelvett \f@ntkey=7 }
  \def\caps{\fam\cpfam \twelvecp \f@ntkey=8 }
  \h@big=10.2\p@{} \h@Big=13.8\p@{} \h@bigg=17.4\p@{} \h@Bigg=21.0\p@{}
  \setbox\strutbox=\hbox{\vrule height 10pt depth 4pt width\z@}
  \samef@nt}
\def\fourteenpoint{\relax
  \textfont0=\fourteenrm \scriptfont0=\tenrm \scriptscriptfont0=\sevenrm
  \textfont1=\fourteeni \scriptfont1=\teni \scriptscriptfont1=\seveni
  \textfont2=\fourteensy \scriptfont2=\tensy \scriptscriptfont2=\sevensy
  \textfont3=\fourteenex \scriptfont3=\fourteenex
    \scriptscriptfont3=\fourteenex
  \textfont\itfam=\fourteenit
  \textfont\slfam=\fourteensl \scriptfont\slfam=\tensl
  \textfont\bffam=\fourteenbf \scriptfont\bffam=\tenbf
    \scriptscriptfont\bffam=\sevenbf
  \textfont\ttfam=\twelvett
  \textfont\cpfam=\twelvecp
  \def\rm{\fam0 \fourteenrm \f@ntkey=0 }
  \def\oldstyle{\fam1 \fourteeni \f@ntkey=1 }
  \def\it{\fam\itfam \fourteenit \f@ntkey=4 }
  \def\sl{\fam\slfam \fourteensl \f@ntkey=5 }
  \def\bf{\fam\bffam \fourteenbf \f@ntkey=6 }
  \def\tt{\fam\ttfam \fourteentt \f@ntkey=7 }
  \def\caps{\fam\cpfam \fourteencp \f@ntkey=8 }
  \h@big=11.9\p@{} \h@Big=16.1\p@{} \h@bigg=20.3\p@{} \h@Bigg=24.5\p@{}
  \setbox\strutbox=\hbox{\vrule height 12pt depth 5pt width\z@}
  \samef@nt}
\def\seventeenpoint{\relax
  \textfont0=\seventeenrm \scriptfont0=\tenrm \scriptscriptfont0=\sevenrm
  \textfont1=\seventeeni \scriptfont1=\teni \scriptscriptfont1=\seveni
  \textfont2=\seventeensy \scriptfont2=\tensy \scriptscriptfont2=\sevensy
  \textfont3=\seventeenex \scriptfont3=\seventeenex
    \scriptscriptfont3=\seventeenex
  \textfont\itfam=\seventeenit
  \textfont\slfam=\seventeensl \scriptfont\slfam=\tensl
  \textfont\bffam=\seventeenbf \scriptfont\bffam=\tenbf
    \scriptscriptfont\bffam=\sevenbf
  \textfont\ttfam=\twelvett
  \textfont\cpfam=\twelvecp
  \def\rm{\fam0 \seventeenrm \f@ntkey=0 }
  \def\oldstyle{\fam1 \seventeeni \f@ntkey=1 }
  \def\it{\fam\itfam \seventeenit \f@ntkey=4 }
  \def\sl{\fam\slfam \seventeensl \f@ntkey=5 }
  \def\bf{\fam\bffam \seventeenbf \f@ntkey=6 }
  \def\tt{\fam\ttfam \seventeentt \f@ntkey=7 }
  \def\caps{\fam\cpfam \seventeencp \f@ntkey=8 }
  \h@big=14.5\p@{} \h@Big=19.5\p@{} \h@bigg=24.7\p@{} \h@Bigg=29.8\p@{}
  \setbox\strutbox=\hbox{\vrule height 14pt depth 6pt width\z@}
  \samef@nt}
\def\twentypoint{\relax
  \textfont0=\twentyrm \scriptfont0=\tenrm \scriptscriptfont0=\sevenrm
  \textfont1=\twentyi \scriptfont1=\teni \scriptscriptfont1=\seveni
  \textfont2=\twentysy \scriptfont2=\tensy \scriptscriptfont2=\sevensy
  \textfont3=\twentyex \scriptfont3=\twentyex
    \scriptscriptfont3=\twentyex
  \textfont\itfam=\twentyit
  \textfont\slfam=\twentysl \scriptfont\slfam=\tensl
  \textfont\bffam=\fourteenbf \scriptfont\bffam=\tenbf
    \scriptscriptfont\bffam=\sevenbf
  \textfont\ttfam=\twelvett
  \textfont\cpfam=\twelvecp
  \def\rm{\fam0 \fourteenrm \f@ntkey=0 }
  \def\oldstyle{\fam1 \twentyi \f@ntkey=1 }
  \def\it{\fam\itfam \twentyit \f@ntkey=4 }
  \def\sl{\fam\slfam \twentysl \f@ntkey=5 }
  \def\bf{\fam\bffam \twentybf \f@ntkey=6 }
  \def\tt{\fam\ttfam \twentytt \f@ntkey=7 }
  \def\caps{\fam\cpfam \twentycp \f@ntkey=8 }
  \h@big=17.0\p@{} \h@Big=23.0\p@{} \h@bigg=29.0\p@{} \h@Bigg=35.0\p@{}
  \setbox\strutbox=\hbox{\vrule height 20pt depth 7pt width\z@}
\samef@nt}
%
%
%
\newdimen\h@big
\newdimen\h@Big
\newdimen\h@bigg
\newdimen\h@Bigg
\def\big#1{{\hbox{$\left#1\vbox to\h@big{}\right.\n@space$}}}
\def\Big#1{{\hbox{$\left#1\vbox to\h@Big{}\right.\n@space$}}}
\def\bigg#1{{\hbox{$\left#1\vbox to\h@bigg{}\right.\n@space$}}}
\def\Bigg#1{{\hbox{$\left#1\vbox to\h@Bigg{}\right.\n@space$}}}
%
%
%
\newskip\normaldisplayskip
\newskip\normaldispshortskip
\newskip\normalparskip
\newskip\normalskipamount
\normalbaselineskip = 20pt plus 0.2pt minus 0.1pt
\normallineskip = 1.5pt plus 0.1pt minus 0.1pt
\normallineskiplimit = 1.5pt
\normaldisplayskip = 20pt plus 5pt minus 10pt
\normaldispshortskip = 6pt plus 5pt
\normalparskip = 6pt plus 2pt minus 1pt
\normalskipamount = 5pt plus 2pt minus 1.5pt
%
%
\def\sp@cing#1{%
  \baselineskip=\normalbaselineskip%
    \multiply\baselineskip by #1%
    \divide\baselineskip by 12%
  \lineskip=\normallineskip%
    \multiply\lineskip by #1%
    \divide\lineskip by 12%
  \lineskiplimit=\normallineskiplimit%
    \multiply\lineskiplimit by #1%
    \divide\lineskiplimit by 12%
  \parskip=\normalparskip%
    \multiply\parskip by #1%
    \divide\parskip by 12%
  \abovedisplayskip=\normaldisplayskip%
    \multiply\abovedisplayskip by #1%
    \divide\abovedisplayskip by 12%
  \belowdisplayskip=\abovedisplayskip%
  \abovedisplayshortskip=\normaldispshortskip%
    \multiply\abovedisplayshortskip by #1%
    \divide\abovedisplayshortskip by 12%
  \belowdisplayshortskip=\abovedisplayshortskip%
    \advance\belowdisplayshortskip by \belowdisplayskip%
    \divide\belowdisplayshortskip by 2%
  \smallskipamount=\normalskipamount%
    \multiply\smallskipamount by #1%
    \divide\smallskipamount by 12%
  \medskipamount=\smallskipamount%
    \multiply\medskipamount by 2%
  \bigskipamount=\smallskipamount%
    \multiply\bigskipamount by 4}
%
%
%
\newcount\fontsize
\def\Eightpoint{\eightpoint\fontsize=8\sp@cing{8}}
\def\Ninepoint{\ninepoint\fontsize=9\sp@cing{9}}
\def\Tenpoint{\tenpoint\fontsize=10\sp@cing{10}}
\def\Twelvepoint{\twelvepoint\fontsize=12\sp@cing{12}}
\def\Fourteenpoint{\fourteenpoint\fontsize=14\sp@cing{14}}
\def\Seventeenpoint{\seventeenpoint\fontsize=14\sp@cing{17}}
\def\Twentypoint{\twentypoint\fontsize=14\sp@cing{20}}
\newcount\spacesize
\def\singlespace{\spacesize=\fontsize
  \multiply\spacesize by 9\divide\spacesize by 12%
  \sp@cing{\spacesize}}
\def\normalspace{\sp@cing{\fontsize}}
\def\doublespace{\spacesize=\fontsize
  \multiply\spacesize by 15\divide\spacesize by 12%
  \sp@cing{\spacesize}}
\Twelvepoint 
\interlinepenalty=50
\interfootnotelinepenalty=5000
\predisplaypenalty=9000
\postdisplaypenalty=500
\hfuzz=1pt
\vfuzz=0.2pt
\catcode`@=12 

\nopagenumbers
\hsize=14.8 cm
\vsize=21.9 cm
\hoffset=6.5mm
\parskip=0mm 
\def\pil{\vphantom{\Big|}} 
\line{}
\vskip 2mm

\Twentypoint 

{\centerline{\bf Dressing the Scalar Glueball} }
\vskip 4mm

{\centerline{M. Boglione and M.R. Pennington}}
\vskip 8mm

\Tenpoint
\baselineskip=4.5mm   
\centerline{Centre for Particle Theory,
University of Durham,}

\centerline{Durham DH1 3LE, U.K.}
\vskip 10mm

{\leftskip 1cm\rightskip 1cm{\noindent {\bf Abstract:} The hadronic dressing of the ten lightest
scalar mesons
is discussed.}\par}
\vskip 8mm

\Twelvepoint
\baselineskip=5mm
\parskip=0mm
Let us imagine that we live in a world without quarks.  QCD suggests
there will still be a spectrum of hadrons, bound states of two or more gluons,
 the lightest of these being stable.  Though this world is far 
 from that explored by experiment,
it is the world of the  lattice calculator.  There the lightest state is a scalar in
the region of 1--2 GeV, just where we expect the 
lowest $q{\overline q}$ scalars
to be.

  The lattice makes definite predictions for the bare glueball mass,
depending on the calculational scheme~: 1550 MeV from UKQCD~(1), 1600 MeV from
Morningstar and Peardon~(2) and 1740 MeV from the GF-11 group at IBM~(3).  Though this
 state is stable in the quenched world of the lattice, the IBM group have pioneered the calculation of its coupling to 2 pseudoscalar
 sources and found this would correspond to a width of some 100 MeV~(4).  Naively,
 one would expect the bare glueball to couple as an $SU(3)_F$ singlet. The
IBM group have calculated that in fact coupling to heavier pseudoscalars
 should be favoured~(4).  Thus the lattice makes
predictions for the scalar gluestate with some element of
 choice about its
exact mass and coupling scheme.  Here we will focus on just one for ease of
presentation.  We will take a mass of 1600 MeV with a coupling pattern as
given by the IBM group (note that this pattern is not very 
sensitive to the exact glueball mass).
However, very similar results apply for other choices.

  Of course, the bare non-decaying state is not what any experiment can observe.  A bare state,
whether composed of glue or quarks, has to be dressed by interactions with mesons in
order to decay. Thus, though we like to think of the $\rho$ or $\phi$ as
$q{\overline q}$ states, their Fock space contains important
$\pi\pi$ and $K{\overline K}$ components, respectively, through which these 
hadrons decay.  For most mesons,
and in particular, the vectors and tensors, these hadronic dressings make rather little difference to the states.
So the hadrons we observe and the underlying $q{\overline q}$ bound states are
 simply related and readily identifiable one from the other.  
 However, as has been emphasised by Tornqvist~(5),
this is not the case for scalars.  This is because their couplings are 
larger and because, in as much as decays to two pseudoscalars dominate, their interactions are $S$--wave
making the opening of the corresponding thresholds especially important,
as Bugg~(6) has long stressed.

  While for vectors and tensors, decays can be reliably modelled by the $^3P_0$
mechanism, this is known to fail for scalars~(7,8).  The Schwinger-Dyson equation for the scalar propagators is
the natural vehicle for calculating these necessarily non-perturbative effects.
In the self-energy loops (Fig.~1) one must sum over all hadronic intermediate states, of these two pseudoscalars are
the most important and the ones we discuss first~(9,10).
\vskip 3.8cm

\Tenpoint
\baselineskip=4.1mm
\noindent{\bf FIGURE 1.} The bare bound state propagator is dressed by hadronic 
interactions.  The dot signifies the dressed hadron propagator. The wiggly
lines on the particles in the loop is to emphasize these too are bound states.

\vskip 2mm
\Twelvepoint
\baselineskip=5mm
\parskip=0mm
  We start with the hadronic interactions turned off and consider the ten
lightest scalars~: the bare glueball and an ideally mixed $q{\overline q}$ nonet.  While
the lattice provides a definite statement about the mass and coupling of the gluestate, we have no such prediction for the
bare $q{\overline q}$ nonet.  We do know that the strange quark adds $\sim 100$ MeV to the mass, so we have just 2 parameters,
the central mass of the $q{\overline q}$ multiplet and its coupling strength to
two pseudoscalars.  These will be fixed by the observed hadron spectrum as we shall see.  By
imposing the constraints of chiral symmetry on the scalar-2 pseudoscalar coupling and
by assuming it to have a form-factor reflecting a roughly common spatial
extent of 0.7~fm for mesons, we can compute, using the Schwinger-Dyson equation,
the effect of dressing  on the 10 bare states.

  We begin with the $I=1/2$ sector, where the LASS experiment~(11,12) has pinned down
the $K^*_0$ parameters close to those given in Table 1.  To get these right, the bare
$n{\overline s}$ state is at 1520 MeV.  The large $K\pi$ and $K\eta'$ couplings
 provide the decays of the dressed hadron and shifts its mass to 1420 MeV.
Though this result is very similar to that of Tornqvist~(9), our calculation differs in an important respect.  Here only
propagators are computed, these determine the right hand cut structure of
the corresponding hadronic amplitudes --- the $D$--function in $N/D$.
In contrast, Tornqvist enforces a much more restrictive range of parameters
by requiring that the $N$--function  is wholly real and fitting data
on the hadronic scattering amplitudes.  This is a drastic oversimplification.  It assumes that amplitudes contain only $s$-channel
dynamics.  Of course, in reality the $N$--function has a left hand cut, the structure
of which is determined by crossed channel dynamics that Tornqvist simply ignores.

  With the parameters of our calculation essentially determined by the $K^*_0$--pole
position, we turn to the $I=1$ sector.  The bare $n{\overline n}$ state is at 1420 MeV,
but the very strong dressing by the $\pi\eta$ and $K{\overline K}$ channels draws the hadron pole towards $K{\overline K}$ threshold
generating an $a_0(980)$ with a Fock space containing only 20\%
 $n{\overline n}$,
but 70\% $K{\overline K}$.  The $I=0$ sector has 3 bare states~:
$n{\overline n}$, $s{\overline s}$ and $gg$.  These mix through their common
hadronic channels $\pi\pi$, $K{\overline K}$, $\eta\eta$, $\eta\eta'$ and $\eta'\eta'$.
The physical scalars are the eigenstates of the resulting mass matrix.
Again states are markedly shifted~: the $f_0(980)$
is drawn to $K{\overline K}$ threshold ---
see Table 1.  Importantly, just as noted by Tornqvist~(9), 
the $a_0(980)$ and the $f_0(980)$
are automatically the dressed partners of the $K_0^*(1430)$.
This is in stark contrast to the model of Janssen {\it et al.}~(13) where the $a_0$ and $f_0$ 
are quite unconnected or the first order perturbative mixing scheme of Amsler
and Close~(14) or Weingarten~(15) where the $a_0(980)$ and $f_0(980)$ 
have to be additional states totally
unrelated to the $K_0^*(1430)$.  Our broad isoscalar is the $f_0(1300)$~(12) 
(Table~1).
In contrast, Roos and Tornqvist~(16) claim a $\sigma(550)$ in their more restrictive scheme, which
as already mentioned violates crossing symmetry~(17).  Our third $f_0$ is closely
related to the bare gluestate.  A common outcome of the bare coupling schemes we have considered
is that the mixing suppresses this hadron's couplings to 2 pseudoscalars reducing its bare 
100 MeV pseudoscalar {\it width} to 25 MeV or so~(10).  Whether the bare gluestate is at 
1600 or 1740 MeV, the hadron's branching fraction to $K{\overline K}$
is always reduced and that to $\eta\eta$ is strongest~(10).  This makes identification with the
$f_J(1710)$ MeV rather unlikely~(18,19,12).
\vskip 2mm
\Tenpoint
\baselineskip=4.5mm
{\leftskip 5.95mm\rightskip 5.95mm{\noindent {\bf TABLE 1.} Masses and
 widths of the ten lightest scalars 
given by their pole positions in MeV and their branching ratios
 to 2 light pseudoscalars from the calculation of Ref. 10.}\par}
\vskip 2mm

\line{\hfil
\vbox{\offinterlineskip
\hrule
\halign{&#&\strut\pil\qquad~~~\hfil#\qquad~~~\cr
&\omit&&\omit&&\omit&&\omit&\cr
& Resonance &&  $m_{pole}$ && $\Gamma_{pole}$ && ~~BR(PS) & \cr
&\omit&&\omit&&\omit&&\omit&\cr
\noalign{\hrule}
&\omit&&\omit&&\omit&&\omit&\cr
& $K_0^*(1430)$ && 1445 && 334~ && 100\%~~ & \cr
& $\;a_0(980)$    &&   1082 && 309~ && 100\%~~ & \cr
& $\;f_0(980)$    &&   1006 &&  54~ && 100\%~~ & \cr  
& $\;f_0(1300)$   &&   1203 && 361~ && 100\%~~ & \cr
& $f_0(1550)^\dagger$   &&   1564 && 108~ && 23\%~~ & \cr
&\omit&&\omit&&\omit&&\omit&\cr}\hrule}\hfil}
{\leftskip 1cm\rightskip 1cm{\noindent $^\dagger$ Here the $f_0(1550)$ results from a 
bare gluestate chosen to be at 1600 MeV~(2,10).}\par}

\Twelvepoint
\baselineskip=5mm
   Of course, 2 pseudoscalars are not the only channels, multipion modes are inevitably
important.  As the lattice gives no idea of the strength of these, we fix the bare gluestate
coupling to these wholly phenomenologically by requiring an output hadron with
a total width of $\sim 100$ MeV, where as before 25 MeV or less is in
2 pseudoscalar channels.  The larger bare coupling shifts the mass a little more, so that a bare 
mass of 1600 MeV moves down to 1564 MeV (Table 1).  

The identification of the
state predicted here with what is seen in experiment requires a careful analysis of a range
of
production processes, not just ${\overline p}p$ annihilation at rest~(20)
or $J/\psi$ radiative decay~(18).  Only by the
consistent analysis of many channels produced in different ways~(18-23) can one arrive at a meaningful
set of parameters for the tenth scalar.  Such analyses have started~(24,25). 
However, it should be remembered that while experiment
determines the dressed states (these being the poles of the $S$--matrix) with 
reasonable accuracy, the poles of the $K$--matrix, which are the underlying states (Fig.~1),
are not fixed unambiguously. Different solutions with differing numbers of
$K$--matrix poles, but very similar $S$--matrix poles, may describe the data
equally well.  Thus the underlying poles are not uniquely determinable from
data with present accuracy and so the {\it true} bare states are not easily found,
despite~(25).
Hence our predictions for the dressed states is what matters.
With these, we should soon know if the $f_0(1500)$~(20-23,12) is the
predominantly glue candidate or not.
\vskip 4mm

\Fourteenpoint
\centerline{\bf REFERENCES}
\baselineskip=4.5mm

\vskip 4mm 
\Tenpoint
\parskip=0mm
\item{    1. } G.S. Bali {\it et al.} (UKQCD), Phys. Lett.  {\bf B389}, 378
               (1993).

\item{    2. } C. Morningstar and M. Peardon, Nucl. Phys. Proc. Suppl. {\bf 53}, 917 (1997). 

\item{    3. } J. Sexton {\it et al.}, Phys. Rev. Lett. {\bf 75}, 4563
               (1995).

\item{    4. } J. Sexton {\it et al.}, presented at ``Lattice
                95'', Melbourne, Australia, Jul 1995, hep-lat/9602022.

\item{    5. } N.A. Tornqvist, Acta Phys. Pol. {\bf B16}, 503 (1985).

\item{    6. } D.V. Bugg, Proc. {\it Hadron '89}, Ajaccio (ed. F.
                  Binon {\it et al}), Editions Frontiers, p. 567. 

\item{    7. } P. Geiger and N. Isgur, Phys. Rev.  {\bf D44}, 799 (1991); 
                 {\bf D47}, 5050 (1993).

\item{    8. } E.S. Ackleh {\it et al.}, Phys. Rev. {\bf
                 54}, 6811 (1996).

\item{    9. } N.A. Tornqvist, Z. Phys. {\bf C68}, 647 (1995).

\item{    10.} M. Boglione and M.R. Pennington, Phys. Rev. Lett. {\bf 79}, 1998 (1997).

\item{    11.} D. Aston {\it et al.}, Nucl. Phys. {\bf B296}, 493 (1988).

\item{    12.} R.M. Barnett {\it et al.}, Review of Particle Physics, Phys. Rev. D {\bf 54}, 30 (1996).

\item{    13.} G. Janssen {\it et al.}, Phys Rev. {\bf D52}, 2690 (1995).

\item{    14.} C. Amsler, F.E. Close, Phys. Rev.  {\bf D53}, 295 (1996);
                   Phys.Lett. {\bf B353}, 385 (1995).
                   
\item{    15.} D. Weingarten, Nucl. Phys. Proc. Suppl. {\bf 53} (1997) 232.

\item{    16.} N.A. Tornqvist, M. Roos, Phys. Rev. Lett. {\bf 76}, 1575(1996).

\item{    17.} N. Isgur and J. Speth, Phys. Rev. Lett. {\bf 77}, 2332 (1996);
              M. Harada {\it et al.}, Phys. Rev. Lett. {\bf 78}, 1603 (1997).

\item{    18.} J.E. Augustin {\it et al.}, Z. Phys. {\bf C36}, 369 (1987);
             Phys. Rev. Lett. {\bf 60}, 2238 (1988); R.M. Baltrusaitis 
             {\it et al.}, Phys. Rev.  {\bf D35}, 2077 (1987).
             
\item{    19.} S.J. Lindenbaum {\it et al.}, Phys. Lett. {\bf B274}, 492 (1992). 

\item{    20.} C. Amsler {\it et al.}, Phys. Lett. {\bf B342}, 433 (1995);
               {\bf B353}, 571 (1995); {\bf B355}, 
               425 (1995).

\item{    21.} F. Binon {\it et al.}, Nuovo Cim. {\bf 80}, 363 (1984); 
D. Alde {\it et al.}, Phys. Lett. {\bf B201}, 160 (1988).

\item{    22.} T.A. Armstrong {\it et al.}, Phys. Lett. {\bf B228}, 536 (1989).

\item{    23.} K. Peters and Ch. Strassburger, contributions to these Proceedings.

\item{    24.} E.g. A. Sarantsev, Proc. {\it Hadron '95}, Manchester,
(ed. M. Birse {\it et al.}), World Sci., p. 384. 

\item{    25.} V.V. Anisovich, contribution to these Proceedings.

\bye